\documentclass{appolb}
\usepackage{graphicx}

\newcommand{\mathd}{\mathrm{d}}

\def\Comment#1{}
\newcommand{\bean}{\begin{eqnarray*}}
\newcommand{\eean}{\end{eqnarray*}}

\newcommand{\gapproxeq}{\lower
.7ex\hbox{$\;\stackrel{\textstyle >}{\sim}\;$}}
\newcommand{\lapproxeq}{\lower
.7ex\hbox{$\;\stackrel{\textstyle <}{\sim}\;$}}

\newcommand\lsim{\mathrel{\rlap{\lower4pt\hbox{\hskip1pt$\sim$}}
    \raise1pt\hbox{$<$}}}
\newcommand\gsim{\mathrel{\rlap{\lower4pt\hbox{\hskip1pt$\sim$}}
    \raise1pt\hbox{$>$}}}
\newcommand{\ba}{\begin{array}}
\newcommand{\ea}{\end{array}}
\newcommand{\nn}{\nonumber}

\newcommand{\be}{\begin{equation}}
\newcommand{\ee}{\end{equation}}
\newcommand{\bear}{\begin{eqnarray}}
\newcommand{\eear}{\end{eqnarray}}
\newcommand{\cO}{{\cal O}}

\newcommand{\mA}{\mathcal{A}}

\newcommand{\mF}{\mathcal{F}}

\def\bat{\begin{array}{cc}}

\newcommand{\Frac}[2]{\frac{\displaystyle #1}{\displaystyle #2}}

\begin{document}


\title{Relations between anomalous and even-parity sectors in AdS/QCD
\thanks{
Presented at Light Cone 2012, 8-13 July 2012, Krakow (Poland).
Talk based on Ref.~\cite{Zuo:2012}.
This work is partially supported by the Italian MIUR PRIN 2009,
the Universidad CEU Cardenal Herrera grants PRCEU-UCH15/10 and PRCEU-UCH35/11,
and the MICINN-INFN fund AIC-D-2011-0818. I would like to thank the organizers for the friendly
environment and the interesting scientific discussions.
I also want to thank P. Colangelo, F. De Fazio, S. Nicotri, F. Giannuzzi and F. Zuo for useful comments.
}
}
\author{Juan Jos\'e Sanz-Cillero
\address{Istituto Nazionale di Fisica Nucleare, Sezione di Bari, Italy}
}
\maketitle
\begin{abstract}
We derive the $\cO(p^6)$ Chiral Perturbation Theory
Lagrangian in the massless quark limit for  a class of gravity dual models of
Quantum Chromodynamics with the chiral  symmetry broken through boundary conditions.
The odd $\cO(p^6)$ couplings are related to the $\cO(p^4)$ low-energy constants (LEC's) in the
even-parity sector.  Some combinations
of even $\cO(p^6)$ couplings  are found to be universal and independent of the peculiarities of the model.
These relations turn out to be  the manifestation at low energies of a broader
relation between anomalous and even-parity  amplitudes.

 \end{abstract}




\vspace*{-9cm}
\hspace*{8cm}
{\bf BARI-TH-2012-664}\\
\vspace*{8cm}

\section{Introduction}

The relation between the anomalous and even-parity sectors of  Quantum Chromodynamics (QCD)
was studied in detail in Ref.~\cite{Zuo:2012} within the framework of holographic
QCD~\cite{Son:2003et,Hirn:2005nr,Sakai:2004cn,Sakai:2005yt}.
This work was motivated by the analysis of  Son and Yamamoto~\cite{Son:2010vc}
of holographic models where the chiral symmetry was broken through boundary conditions
(b.c.'s).    An interesting relation was found therein between
the left--right correlator $\Pi_{LR}(Q^2)$
and the transverse part $w_T(Q^2)$ of the anomalous
$AVV$ Green's function~\cite{Son:2010vc}   (studies of such a relation
can be found in~\cite{other-AVV}).
At low energies, this turned into a relation between
the $\cO(p^4)$ even--parity Chiral Perturbation Theory ($\chi$PT)  coupling $L_{10}$
and the $\cO(p^6)$ odd-intrinsic-parity coupling  $C_{22}^W$~\cite{Knecht:2011wh}.  

In~\cite{Zuo:2012}, we derived the remaining $\cO(p^6)$ odd-sector couplings
--in the massless quark limit considered all along the analysis--  and found analogous relations
with the $\cO(p^4)$ even low-energy constants (LEC's).
We  focused on the odd couplings $C_{22}^W$ and $C_{23}^W$~\cite{Bijnens:1999sh},
which can be directly related
to the transition of a pion into  two photons and two axial-vector currents, respectively.
These  amplitudes are found to be related to the vector form factor of the pion
and the axial-vector form factor into three pions~\cite{Zuo:2012}.     

In the kind of models studied in this paper, with the chiral symmetry broken through b.c.'s,
the action is composed by the Yang--Mills (YM) and Chern--Simons (CS) terms, describing
the even and anomalous QCD sectors,
respectively~\cite{Son:2003et,Hirn:2005nr,Sakai:2004cn,Sakai:2005yt}:
\begin{eqnarray}
  S &=& S_{\rm YM}+S_{\rm CS}
  \label{eq.5Daction}
  \\
  \label{eq:YM}
  S_{\rm YM} &=& -\int\! d^5x \mbox{tr}\left[-f^2(z){\cal F}_{z\mu}^2
  + \frac{1}{2g^2(z)}{\cal F}_{\mu\nu}^2 \right],
  \nn\\
  \label{eq:CS}
  S_{\rm CS} &=& -  \, \Frac{N_C}{24\pi^2} \,
  \int\! \mbox{tr}
  \left[{\cal AF}^2+\frac{i}{2}{\cal A}^3{\cal F}-\frac{1}{10}{\cal A}^5
\right]\, .
\nn
\end{eqnarray}
with $N_C$ the number of colors and
the fifth coordinate $z$ running from $-z_0$ to $z_0$ with $0<z_0\le+\infty$.
${\cal A}(x,z)={\cal A}_M dx^M$ is the 5D $U(N_f)$ gauge field and
${\cal F}=d{\cal A}-i{\cal A} \wedge {\cal A}$ is the field strength.

Chiral symmetry can be realized
as a 5D gauge symmetry localized on the two boundaries at $z=\pm z_0$.
The gauging of the chiral symmetry allows one to naturally introduce
the corresponding right and left current sources, respectively $r_\mu(x)$ and $\ell_\mu(x)$,
through the ultraviolet b.c.'s    $\mA_\mu(x, -z_0)=\ell_\mu(x)$ and
$\mA_\mu(x,z_0)=r_\mu(x)$~\cite{Zuo:2012}.
In the 5D gauge $\mA_z=0$, one has the   field   decomposition  in on-shell states,
\begin{equation}
{\cal A}_\mu(x,z)=i\Gamma_\mu(x)+\frac{u_\mu(x)}{2}\psi_0(z)
+\sum_{n=1}^\infty v_\mu^n(x)\psi_{2n-1}(z)
+\sum_{n=1}^\infty a_\mu^n(x)\psi_{2n}(z)\, ,
\label{eq.Amu-decomposition2}
\end{equation}
where the commonly used tensors $u_\mu(x)$ and $\Gamma_\mu(x)$  from
$\chi$PT contain the chiral Goldstones and $\ell_\mu(x)$
and $r_\mu(x)$~\cite{Bijnens:1999sh,Ecker:1988te}.
The resonance wave-functions $\psi_n(z)$ are provided by
the normalizable eigenfunctions of the equation of motion (EoM)  for the transverse part of the gauge field
and the pion wave-function $\psi_0(z)$ is the solution of the EoM at $q^2 = 0$ with b.c.'s
$\psi_0(\pm z_0) = \pm 1$.

Once we have rewritten the 5D fields in terms of the chiral Goldstones and
vector and axial-vector resonances, the derivation of the meson Lagrangian
is  straightforward.
We substitute the   $\mA_\mu$  decomposition
provided in eq.~(\ref{eq.Amu-decomposition2})  in the 5D action~(\ref{eq.5Daction}).
%
%
This yields  terms without resonance  fields~\cite{Hirn:2005nr,Sakai:2004cn,Sakai:2005yt} which,
at low energies, provide the Wess-Zumino-Witten (WZW)  Lagrangian
($\cO(p^4)$)~\cite{Sakai:2004cn,Sakai:2005yt}
and the  $\cO(p^2)$ and $\cO(p^4)$  $\chi$PT action~\cite{Gasser:1983yg}   
with the corresponding LEC's given in terms of  the corresponding
5D integrals of pion wave-function~\cite{Hirn:2005nr}.

\section{Sum-rules and chiral couplings at $\cO(p^6)$}
\label{sec.op6-LECs}


The $\cO(p^6)$ LECs are generated by the intermediate  resonance exchanges.
More precisely, we need just the one-resonance exchanges,
given by the terms of the action with one resonance field.  
The couplings $a_{Vv^n}$, $a_{Aa^n}$, $b_{v^n\pi\pi}$, $b_{a^n\pi^3}$...
are defined by the corresponding 5D integrals of the wave-functions.    
At the level of the generating functional, in order to compute the diagrams with intermediate
resonances one must perform the functional integration over the heavy resonance
configurations in the low-energy limit~\cite{Zuo:2012,Ecker:1988te}.

Before proceeding to the actual computation,   we will consider a series of resonance sum-rules
which will be needed for the extraction  of some  LEC's and the amplitudes.
They are obtained through the EoM's of the 5D fields
and the completeness condition for the wave-function solutions $\psi_m(z)$~\cite{Zuo:2012}.
We obtain, for instance,
\begin{eqnarray}
&&\sum_{n=1}^\infty a_{Vv^n} b_{v^n\pi\pi} m_{v^n}^2=2f_\pi^2,\,\,~\qquad
\sum_{n=1}^\infty a_{Vv^n} b_{v^n\pi\pi}=4L_9\, ,
\label{eq.sum-rule-VFF}
\\
&&\sum_{n=1}^\infty 3 a_{Aa^n} c_{a^n} =2\, ,\,\,~\qquad
\sum_{n=1}^\infty 3 a_{Aa^n}c_{a^n}/m_{a^n}^2=4(L_9-8L_1)/f_\pi^2\, .
\label{eq.sum-rule-AFF}
\end{eqnarray}
respectively related to  the $\pi\pi$ vector form factor~(VFF)   
and  the $\pi\pi\pi$ axial-vector form factor~(AFF) at high energies.

Through the integration of the heavy resonances in the generating functional
we obtain all the odd-parity sector LEC's   $C_k^W$  in the massless quark case.
By means of wave-function completeness relations and EoM's they can be   reexpressed
in terms of the $\cO(p^4)$ $\chi$PT couplings of the even sector $L_1$ and $L_9$
and a constant $Z$~\cite{Zuo:2012}.
In particular, we find
\begin{eqnarray}
C_{22}^W  =  \frac{N_C}{32\pi^2f_\pi^2}L_9\label{eq:C22}
\, , \qquad\qquad
C_{23}^W =  \frac{N_C}{96\pi^2f_\pi^2}(L_9-8L_1).
\label{eq:C23}
\label{eq.CWk-Lj-rel}
\end{eqnarray}
Taking into account the relation  $L_9=-L_{10}$ in this kind of holographic models
~\cite{Hirn:2005nr},  
we recover the result $C_{22}^W= - \frac{N_C}{32\pi^2 f_\pi^2} L_{10}$~\cite{Knecht:2011wh}.

It is also possible to compute the $\cO(p^6)$ even-sector LEC's in the holographic model.
In particular, we find some relations  independent of the peculiarities of the model:
the   $\gamma\gamma\to \pi^0\pi^0$ amplitude is ruled by~\cite{Zuo:2012,Gasser:2005ud}
\begin{eqnarray}
&&
256\pi^4 ~ f_\pi^2~(8 C_{53} + 8 C_{55} +C_{56}+C_{57}+2 C_{59})
= N_C^2 \,,\nn\\
&&
-128\pi^4~ f_\pi^2~(  C_{56}+C_{57}+2 C_{59} )= N_C^2/6 \, ,
\end{eqnarray}
and for $\gamma\gamma\to \pi^+\pi^-$~\cite{Zuo:2012,Gasser:2006qa},
\begin{eqnarray}
&&
256\pi^4~ f_\pi^2~ (8 C_{53} - 8 C_{55} +C_{56}+C_{57}
-2 C_{59}+4 C_{78} +8 C_{87}-4 C_{88})=0\,,\nn\\
&&
-128\pi^4~ f_\pi^2~( C_{56}+C_{57}-2 C_{59}-4 C_{78} )=0\,.
\end{eqnarray}

\section{Relations between anomalous  and even-parity  amplitudes}
\label{sec.amplitude-rel}

\subsection{Green's function relation:   LR versus  AVV correlator}

The relations between the odd-sector $\cO(p^6)$ constants and the $\cO(p^4)$   constants
in the even sector indicate possible relations between hadronic amplitudes.
One example is the Son-Yamamoto relation between the transverse
structure function $w_T(Q^2)$ of the $AVV$ Green's function and the left-right correlator
$\Pi_{LR}(Q^2)$~\cite{Son:2010vc}:
\begin{equation}
w_T(Q^2)=\frac{N_C}{Q^2} + \frac{N_C}{f_\pi^2}\, \Pi_{LR}(Q^2)  \, ,\label{eq-SY}
\end{equation}
where the Euclidean squared momentum transfer $Q^2=-q^2$.
Taking the $Q^2\to 0$ limit on both sides, one gets the $C_{22}^W$
LEC relation but with $L_{9}=-L_{10}$~\cite{Knecht:2011wh}.


\subsection{Form factor relation:  $\gamma^*\to\pi\pi$ versus $\pi \to \gamma\gamma^*$}

Other studies   in two specific models showed that the $\pi^0\to\gamma\gamma^*$ transition
form-factor $\mF_{\pi\gamma^*\gamma^*}(Q^2,0)$
was equal to the pion vector form-factor $\mF_{\gamma^*\pi\pi}(Q^2)$ up to
normalization~\cite{Grigoryan:2008cc,Stoffers:2011xe}.
This  relation was found to be universal in the class of models
considered here~\cite{Zuo:2012}:
\begin{equation}
\mF_{\pi\gamma\gamma}(Q^2,0)=\frac{N_C}{12\pi^2f_\pi} \,
\mF_{\gamma\pi\pi}(Q^2)
\, .
\label{eq-VFFTFF}
\end{equation}
In the low-energy limit   $Q^2\to 0$  we recover the  $C_{22}^W$ relation in~(\ref{eq:C22}).

\begin{figure}[ht]
\centering
	\includegraphics[width=0.49\textwidth]{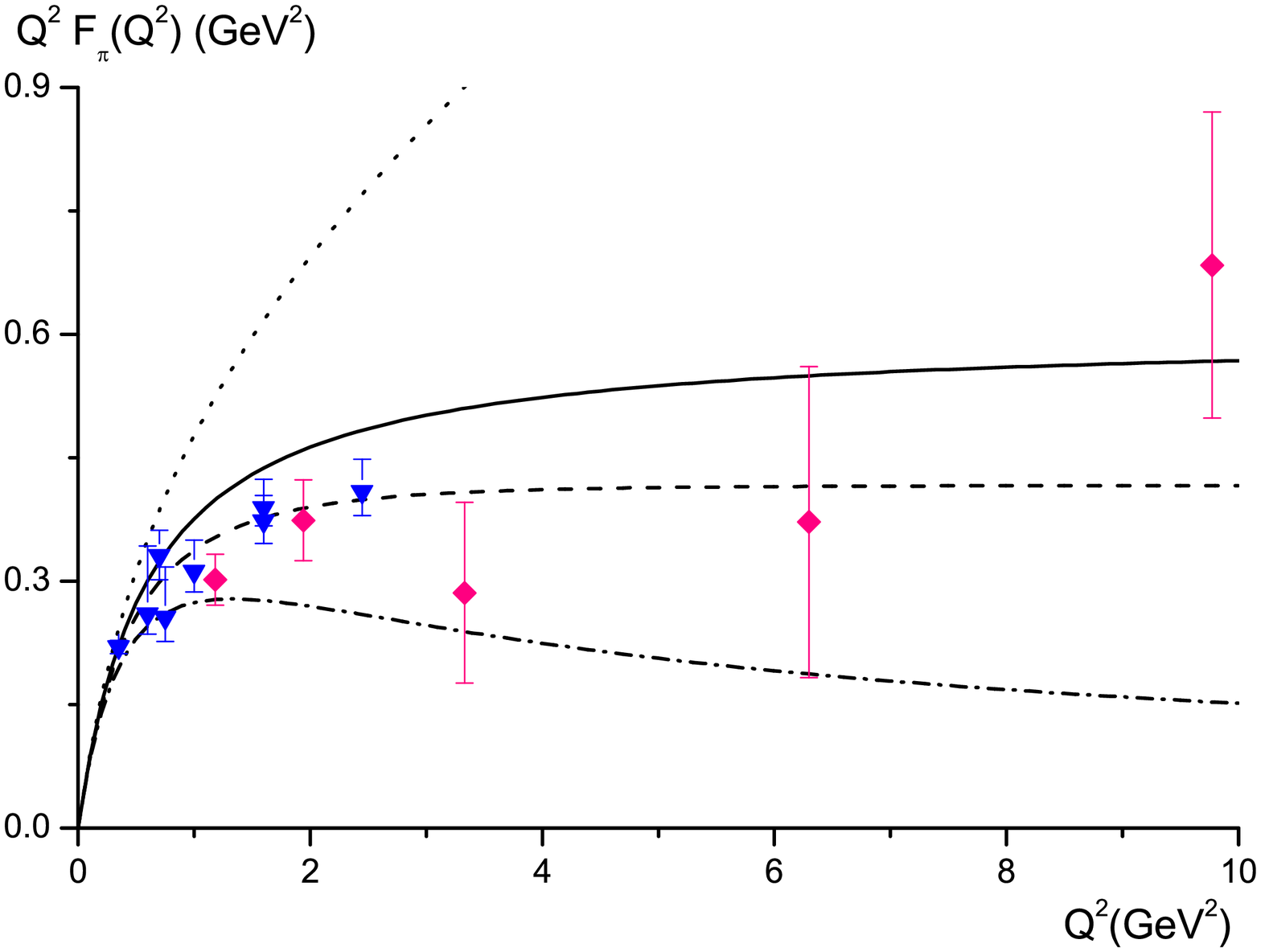}
	\includegraphics[width=0.49\textwidth]{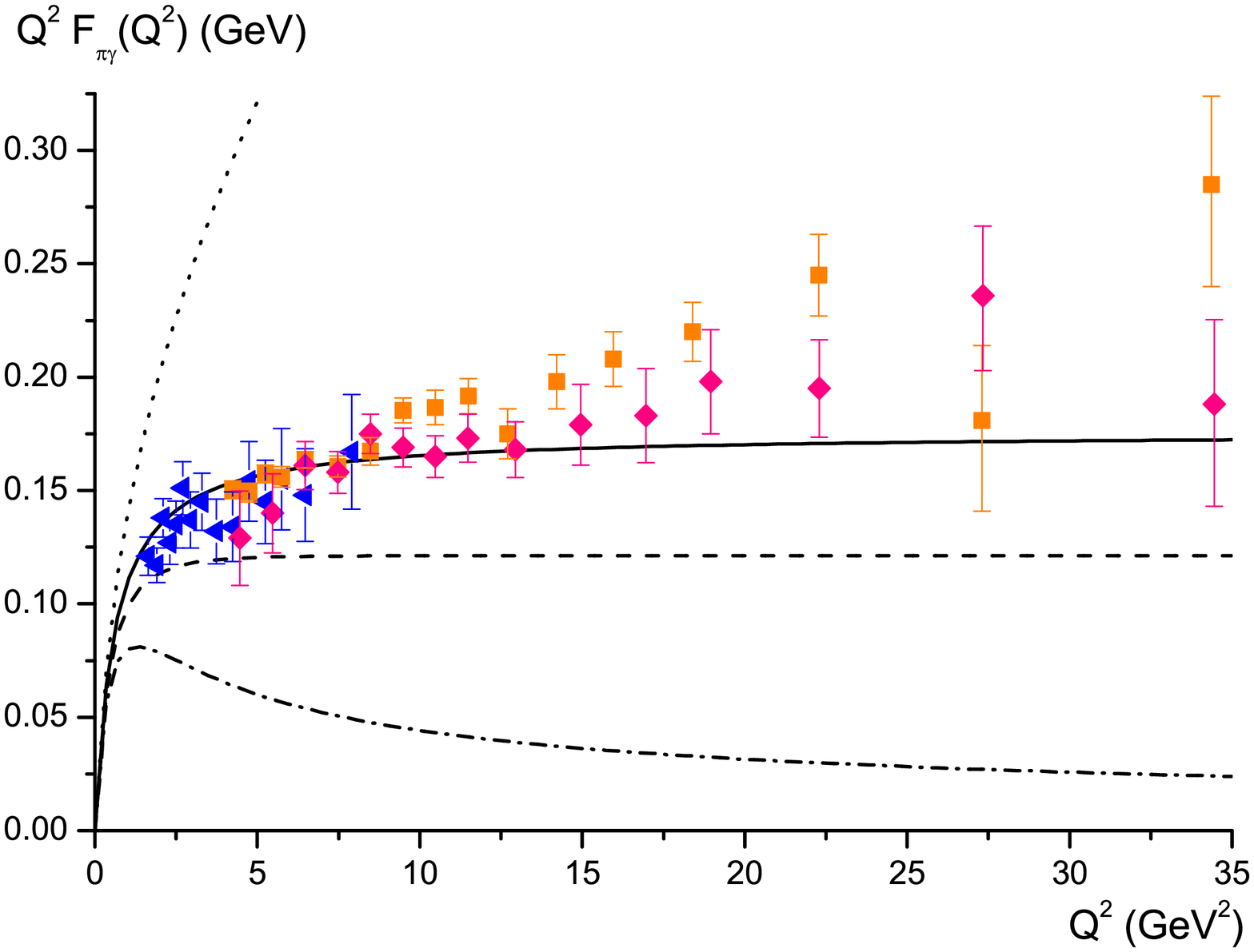}
\caption{\it a) Vector form factor $\mF_{\gamma^*\pi\pi}(Q^2)$ from the flat~\cite{Son:2003et},
``Cosh''~\cite{Son:2003et},
hard wall~\cite{Hirn:2005nr} and Sakai-Sugimoto~\cite{Sakai:2004cn,Sakai:2005yt} models,
denoted by the dotted, solid, dashed and dash-dotted lines, respectively.
The experimental data is taken
 from~\cite{Bebek:1977pe} (diamonds) and~\cite{Huber:2008id} (triangles).
b) Anomalous $\pi\gamma\gamma^*$ form factor:
experimental data  from CLEO~\cite{Gronberg:1997fj}
(triangles),  BABAR~\cite{Aubert:2009mc} (squares) and BELLE
collaboration~\cite{Uehara:2012ag} (diamonds).
 }
 \label{fig:FF}
\end{figure}

The form-factors  from different models may have completely different behaviours
at $Q^2\to \infty$~\cite{Zuo:2011sk},
as one can see in Fig.~\ref{fig:FF} compared to the experimental data.
In order to reproduce the observed $1/Q^2$
behavior for the form factors, the models need to be asymptotically AdS in the UV,
as this is the case of the  ``cosh'' and hard wall models~\cite{Son:2003et,Hirn:2005nr}.
one finds the large-$Q^2$ behavior of the
form factors~\cite{Grigoryan:2008cc}
\begin{equation}
\mF_{\pi \gamma^*\gamma^*}(Q^2,0)
\quad \stackrel{Q^2\to\infty}{\longrightarrow} \quad \Frac{N_C g_5^2 f_\pi}{12\pi^2 Q^2}\, .
\end{equation}
If the 5D coupling $g_5$ is fixed by the perturbative QCD
logarithmic term of the axial-vector correlator at short distances, i.e.,
$g_5^2=24\pi^2/N_C$~\cite{Son:2003et},
one recovers the  asymptotic behavior
$Q^2 \mF_{\pi \gamma^*\gamma^*}(Q^2,0) \stackrel{Q^2\to\infty}{\longrightarrow}
2 f_\pi$~\cite{Lepage:1979zb}.        

\subsection{Form factor relation: $A\to \pi\pi\pi$ versus $\pi\to AA$}

A similar relation shows up between form factors
involving the axial-vector source,
related to the $C_{23}^W$ expression  in~(\ref{eq:C23}).
This $\cO(p^6)$ odd-parity coupling   
is related to the $\pi\to AA$ transition form-factor~\cite{Zuo:2012},
\begin{eqnarray}
&&\int<\pi^c(p)|T\{j_\mu^{5a}(x)j_\nu^{5b}(0)\}|0>e^{-iq_1x} \mathd^4x
\nonumber\\
&&~~
\qquad\qquad
=\frac{i~N_C}{24\pi^2f_\pi}~d^{abc}~\epsilon_{\mu\nu\alpha\beta}~q^{\alpha}_1
q^{\beta}_2~\mF_{\pi AA}(Q_1^2,Q_2^2),
\end{eqnarray}
with $p=q_1+q_2$, $Q_1^2=-q_1^2$, $Q_2^2=-q_2^2$.
In Ref.~\cite{Zuo:2012} we found that it is possible to relate this amplitude
to the $A\to\pi\pi\pi$ form-factor in the massless quark limit~\cite{Dumm:2009va},
\begin{eqnarray}
&&<\pi^a(p_1)\pi^b(p_2)\pi^c(p_3)|i j_\mu^{5d}|0>
=f^{bce}f^{ade}~P_\mu^{\nu\perp}(q)
\\
&&\qquad \qquad\times [\mF_1(Q^2,s,t)(p_1-p_3)_\nu
+\mF_1(Q^2,t,s)(p_2-p_3)_\nu]+(a\leftrightarrow c)  \, ,
\nn
\end{eqnarray}
with $q=p_1+p_2+p_3$, $P_\alpha^{\mu\perp}(q)=\eta_\alpha^\mu-q_\alpha q^\mu/q^2$,
$Q^2=-q^2$, $s=(p_1+p_3)^2$, $t=(p_2+p_3)^2$,  $u=(p_1+p_2)^2$
and the axial-vector current $j_\mu^{5 a}$.

In the particular  kinematical regime $s=t=0$ we find the relation
between anomalous and even-parity sectors~\cite{Zuo:2012},
\be
\Frac{3 f_\pi}{2}\, \mF_{1}(Q^2,0,0) \,\,=\,\,   \, \mF_{\pi AA}(Q^2,0)
\,\,=\,\,
 1\,
-\, \sum_n \Frac{3 a_{A a^n} c_{a^n}}{2}\, \Frac{Q^2}{m_{a^n}^2+Q^2}
\, .
\label{eq.AFF}
\ee
The $C_{23}^W$ relation in~(\ref{eq:C23}) is recovered by means of  the low energy expansion
of $\mF_1(Q^2,0,0)$ and $\mF_{\pi AA}(Q^2,0)$  in ~(\ref{eq.AFF}).

Based on these results,   one may speculate that the relations between
the $\cO(p^6)$ odd and $\cO(p^4)$ even-sector  LEC's
represent  the manifestation at low energies of further amplitude relations.



\end{document}